\documentstyle[twocolumn,aps,epsf]{revtex}
\begin{document}
\draft
\title{
Quasi-Particle Behavior of Composite Fermions in the Half-Filled Landau Level}
%
%\author{Peter Kopietz}
\author{Peter Kopietz$^{1}$ and Guillermo E. Castilla$^{2}$}
\address{
%Institut f\"{u}r Theoretische Physik der Universit\"{a}t G\"{o}ttingen,
%Bunsenstr.9, D-37073 G\"{o}ttingen, Germany}
$^{1}$Institut f\"{u}r Theoretische Physik der Universit\"{a}t G\"{o}ttingen,
Bunsenstr.9, D-37073 G\"{o}ttingen, Germany\\
$^{2}$Department of Physics, University of California at Riverside, CA 92521, USA}

\date{November 8, 1996}
\maketitle
\begin{abstract}
We calculate the effect of infrared fluctuations
of the Chern-Simons gauge field
on the single-particle Green's function of composite fermions
in the half-filled Landau level  
via higher-dimensional bosonization 
on a curved Fermi surface.
We find that composite fermions remain well-defined quasi-particles, with an
effective mass given by the mean-field value, 
but with anomalously large
damping and a spectral function that contains considerable
weight away from the quasi-particle peak.

\end{abstract}
\narrowtext

There exists growing experimental 
evidence\cite{Willett93}
that the fractional quantum Hall state at
Landau level filling $\nu = {1}/{2}$ 
is metallic. However,
the single-particle excitations are not dressed electrons, but
so-called composite fermions\cite{Jain89},
which are born out of the
highly correlated motion of the two-dimensional electron gas
in a strong external magnetic field.  
Composite fermions in the half filled Landau level
can be viewed as spin-polarized electrons that are bound to a flux tube
carrying two flux quanta. 
Theoretically the flux attachment can be modeled 
by coupling the electrons to
a Chern-Simons gauge field\cite{Halperin93}.
At the mean-field level, where fluctuations of the gauge field are
ignored, the magnetic field generated by the Chern-Simons field
exactly cancels the external magnetic field $B$,
so that mean-field theory  predicts that
composite fermions in the half filled Landau level
should behave like free spinless fermions without magnetic field, 
with Fermi wave-vector $k_F = (4 \pi n_e )^{1/2}$. Here $n_e$ is the areal
density of the electron gas.

The existence of a  well-defined Fermi surface has
been confirmed by several experimental techniques, such as
surface acoustic wave-experiments, measurements of Shubnikov-de Haas
oscillations, and magnetic focusing experiments\cite{Willett93}.
Although 
the energy dispersion  of the composite fermions
as function of momentum ${\bf{k}}$ is not known, 
and a number of experiments have found a strong enhancement
of the effective mass $m^{\ast}$\cite{Leadley94},
there exists general agreement that the composite fermions
are well-defined quasi-particles.

Theoretically the situation is less clear. 
In their seminal paper
Halperin, Lee, and Read\cite{Halperin93} 
showed that the leading self-energy
correction due to fluctuations of the Chern-Simons 
field (Eq.(\ref{eq:GW}) below)
completely destroys the quasi-particle peak in the spectral function
predicted by mean-field theory.
There have been various attempts to calculate
non-perturbatively
the effect of gauge-field fluctuations on
the single-particle Green's function\cite{Gan93,Kwon94}, 
but the results are contradictory.
This is possibly due to the gauge-dependence of the
single-particle Green's function; yet, 
we expect physical quantities 
derived from it to be gauge-invariant.
Until now, the experimental fact that composite fermions
manifest themselves as well-defined quasi-particles
could not be justified theoretically.
In this Letter we shall give such a justification,
using our recently developed 
higher-dimensional bosonization approach for curved Fermi surfaces\cite{Kopietz96}.
Specifically,
we show that composite fermions 
remain well-defined quasi-particles
even if the infrared fluctuations of the Chern-Simons gauge field are taken into account.
Remarkably, we also find that these fluctuations
do not lead to a divergence of the effective mass.

\begin{figure}
%\vspace{1cm}
\epsfysize3.3cm 
\hspace{23mm}
\epsfbox{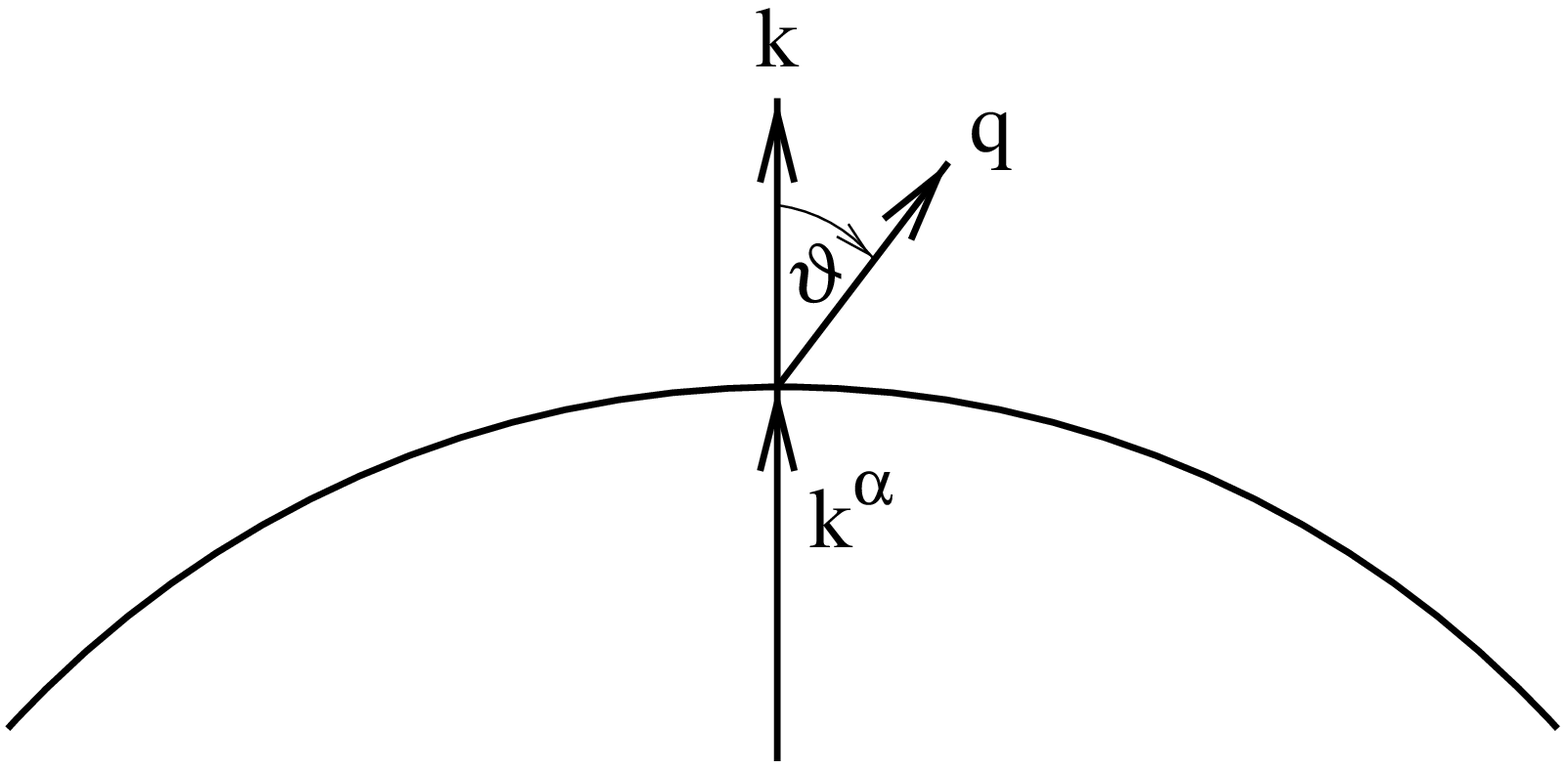}
%\vspace{1cm}
\caption{
Wave-vectors ${\bf{k}}$ and ${\bf{q}}$ measured with
respect to an arbitrary point ${\bf{k}}^{\alpha}$
on the Fermi surface.
}
\label{fig:coordinate}
\end{figure}

To introduce our notation and to set the stage for the
calculations that follow, we first
evaluate the leading self-energy correction
(the so-called GW self energy) due to fluctuations of the
Chern-Simons field in the coordinate system shown in 
Fig.\ref{fig:coordinate}.
As we shall see, such a coordinate system plays a crucial role in our 
bosonization approach.
To leading order in the
RPA (random phase approximation) gauge-field
propagator $h^{{\rm RPA}, \alpha }_{\tilde{q}}$, 
the imaginary frequency self-energy 
of the composite fermions due to fluctuations
of the gauge field can be written as\cite{Halperin93}
 \begin{equation}
 \Sigma_{\rm GW}^{\alpha} ( \tilde{k} ) =
 - \frac{1}{\beta V} \sum_{ \tilde{q} }
 h^{{\rm RPA} , \alpha}_{\tilde{q}} G_0^{\alpha} ( \tilde{k} + \tilde{q} )
 \label{eq:GW}
 \; \; \; ,
 \end{equation}
where $\beta$ is the
inverse temperature, $V$ is the volume of the system,
and $G_0^{\alpha} ( \tilde{k} ) = [ i \tilde{\omega}_n - \xi^{\alpha}_{\bf{k}} ]^{-1} $
is the mean-field Green's function.
We use the notation
$\tilde{k} = [ {\bf{k}} , i \tilde{\omega}_n ]$,
$\tilde{q} = [ {\bf{q}}, i \omega_n]$,
with $\tilde{\omega}_n = 2 \pi (n + 
\frac{1}{2}) / \beta $ and
${\omega}_n = 2 \pi n / \beta$. 
The superscript $\alpha$ indicates that wave-vectors are measured with
respect to ${\bf{k}}^{\alpha}$.
Assuming a spherical Fermi surface, the mean-field energy dispersion
is $\xi^{\alpha}_{\bf{k}} = {\bf{v}}^{\alpha} \cdot {\bf{k}} + 
\frac{ {\bf{k}}^2}{2 m}$, where ${\bf{v}}^{\alpha} = {\bf{k}}^{\alpha} / m$,
and $m$ is the mean-field mass of the composite fermions. 
Throughout this work we shall set $\hbar =1$.
In Coulomb gauge the  propagator of the transverse Chern-Simons field
is at long wave-lengths and for frequencies 
$| \omega_n | 
{ \raisebox{-0.5ex}{$\; \stackrel{<}{\sim} \;$}}
v_F | {\bf{q}} |$
of the form
 \begin{equation}
 h^{{\rm RPA} , \alpha}_{\tilde{q}}
 = -  \frac{2 \pi}{m  } \;
 \frac{ 1 - ({\hat{\bf{k}}}^{\alpha} \cdot {\hat{\bf{q}}} )^2}
 { \left( {|{\bf{q}}| }/{ q_c } \right)^{\eta} + 
  | \omega_{n}| /( v_{F} | {\bf{q}} |)  
  }
 \label{eq:hrpa}
 \; \; \; ,
 \end{equation}
where $v_F = | {\bf{v}}^{\alpha} |$ and
$\hat{\bf{k}}^{\alpha} = {\bf{k}}^{\alpha} / k_F$.
The momentum scale $q_c$ 
and the exponent $\eta$ depend on the nature
of the density-density interaction between the electrons.
For long-range Coulomb forces
$q_c = k_F^2 m / e^2$ and
$\eta = 1$, while $\eta =2$ if the Coulomb interaction
is screened by metal plates. 
For simplicity we shall assume
throughout this work that $1 < \eta \leq 2$, and that $q_c \ll k_F$.
The mean-field mass $m$ should be considered
as an effective mass which takes into account 
Landau level mixing and
short wave-length fluctuations 
not explicitly considered here. 
Simple physical arguments\cite{Halperin93} and 
approximate microscopic calculations\cite{Morf95} yield
$m \propto  ( e^2 \ell)^{-1}  \propto \sqrt{B}$,
where $\ell = \sqrt{  c / ( eB)}$ is the magnetic length.
In this work we are interested in 
the infrared regime.
We shall then impose an ultraviolet cutoff
$\kappa$ on all loop-integrations.
For $\kappa \gg q_c$ 
the infrared physics is independent of $\kappa$.

To perform the integrations in Eq.(\ref{eq:GW}), 
we may restrict
ourselves to external wave-vectors of the form
${\bf{k}} = k_{\|} \hat{\bf{k}}^{\alpha}$, as shown in Fig.\ref{fig:coordinate}.  
The spherical symmetry implies that the  
usual self-energy $\Sigma ( {\bf{k}} , i \tilde{\omega}_n )$ (where
${\bf{k}}$ is measured with respect to the origin of the Fermi sphere)
can obtained by replacing $k_{\|} \rightarrow | {\bf{k}} | - k_F$
in $\Sigma^{\alpha} ( k_{\|} \hat{\bf{k}}^{\alpha} , i \tilde{\omega}_n )$.
The energy dispersion on the right-hand side of Eq.(\ref{eq:GW})
can then be written as
 $\xi^{\alpha}_{ {\bf{k}} + {\bf{q}} } =
 \xi_{ k_{\|}   }
 +  ( 1 + \frac{k_{\|}}{k_F} ) {\bf{v}}^{\alpha} \cdot  {\bf{q}} 
 + \frac{ {\bf{q}}^2 }{2 m}$,
where
 $ \xi_{  k_{\|} }
 = v_F k_{\|} +  k_{\|}^2/(2 m)$.
Using the circular coordinates in
Fig.\ref{fig:coordinate}, the $\vartheta$-integration yields 
 \begin{eqnarray}
 \Sigma_{\rm GW}^{\alpha} ( k_{\|} , i \tilde{\omega}_n )
 & = & \frac{ 1}{2 \pi m ( 1 + \frac{k_{\|}}{k_F})}
 \int_{0}^{\kappa} dq  q
 \int_{ - v_F q}^{v_F q} d \omega
 \nonumber
 \\
 & \times &
 \frac{1}{ | \omega | + \Gamma_q} Z ( W ( q , \omega) )
 \label{eq:GWspherical2}
 \; \; \; ,
 \end{eqnarray}
with
 $
 \Gamma_q = 
  v_F q^{1 + \eta} / q_c^{\eta}$ and
 $Z ( W ) 
 = W - \sqrt{ W^2 -1 }$, where
the root has to be taken such that $| Z | < 1$.
Here 
 \begin{equation}
 W ( q , \omega ) = 
 \frac{G_0^{-1} + i \omega - \frac{ q^2}{2 m} }{v_F q ( 1 + \frac{k_{\|}}{k_F} ) }
 \label{eq:Wdef}
 \; \; \; ,
 \end{equation}
where
 $G_0^{-1} = i \tilde{\omega}_n - \xi_{k_{\|}}$.
For our purpose it is sufficient to know the asymptotic behavior
of $Z ( W)$  for small $W$, which is given by
 $Z ( W ) \sim
 - i {\rm sgn} ( {\rm  Im } W ) + O ( W)$.
Because $ |\omega | / ( v_F q ) 
{ \raisebox{-0.5ex}{$\; \stackrel{<}{\sim} \;$}} 1$ 
and $\frac{q^2 }{  2 m } 
{ \raisebox{-0.5ex}{$\; \stackrel{<}{\sim} \;$}} v_F q$ 
in the domain of integration in Eq.(\ref{eq:GWspherical2}), the
condition $| W | \ll 1$ is equivalent with
$q { \raisebox{-0.5ex}{$\; \stackrel{>}{\sim} \;$}}  k_0$, where 
 $k_0 = {\rm max} \left\{ | k_{\|} | , | \tilde{\omega}_n | / v_F \right\}$.
It is now easy to see that the leading singular behavior
of $\Sigma^{\alpha}_{\rm GW} ( \tilde{k} )$ for small wave-vectors and frequencies
is due to the regime $ q 
{ \raisebox{-0.5ex}{$\; \stackrel{>}{\sim} \;$}} k_0 $ 
and $\omega 
{ \raisebox{-0.5ex}{$\; \stackrel{<}{\sim} \;$}} | \tilde{\omega}_n | $. 
In this regime
we may approximate
 $Z ( W ( q , \omega ) ) \approx - i {\rm sgn} (  \tilde{\omega}_n + \omega )$.
Now the frequency integration in Eq.(\ref{eq:GWspherical2})
can be performed, and the
dependence of the remaining one-dimensional $q$-integral on the external frequency
can be scaled out. 
For convenience we define dimensionless
momenta
$\bar{k}_{\|} = k_{\|} / q_c$ and energies
$\bar{\omega}_n = \tilde{\omega}_n / ( v_F q_c) $.  
If $| \bar{k}_{\|} |$ is not much larger than
$| \bar{\omega}_n |$, it is easy to show that
to leading order\cite{Halperin93,Kim94}
 \begin{equation}
  ( v_F q_c )^{-1} {\Sigma}_{\rm GW}^{\alpha} ( k_{\|} , i \tilde{\omega}_n )
  \sim  
  -    
  i 
  {\rm sgn} ( \tilde{\omega}_n ) 
  | \bar{\omega}_n |^{ \frac{2}{1 + \eta} }
  c_{\eta} g
 \label{eq:GWspherical6}
 \; \; \; ,
 \end{equation}
where $c_{\eta}$ is a positive numerical constant
of the order of unity\cite{Kim94},
and $g = q_c / k_F$.
From Eq.(\ref{eq:GWspherical6}) it is obvious that, after analytic
continuation $ i \tilde{\omega}_n \rightarrow \omega + i 0^{+}$,
${\rm Im} \Sigma^{\alpha}_{\rm GW}$
has the same order of magnitude as 
${\rm Re} \Sigma^{\alpha}_{\rm GW}$, so that
the spectral function does not exhibit a well-defined
quasi-particle peak.
Thus, lowest order perturbation theory suggests that composite fermions
are {\it{not}} well-defined quasi-particles, in contradiction with
the experiments\cite{Willett93}.
If one uses the maximum of the spectral function
to define the energy of the renormalized 
quasi-particle\cite{Halperin93,Kim94} (ignoring the
fact that perturbation theory suggests
that such a quasi-particle really cannot propagate
because the damping is not small), one finds
that the energy of the composite fermion
vanishes as $| \bar{k}_{\|}|^{\frac{1 + \eta}{2}}$ for
$\bar{k}_{\|} \rightarrow 0$, 
indicating a divergence of the effective mass\cite{footnoteeffmess}.
Although an enhanced $m^{\ast}$ has been observed
experimentally\cite{Willett93} (see, however, Ref.\cite{Leadley94}),
the observed enhancement is much stronger than
the above theoretical prediction.

We now use our non-perturbative higher-dimensional bosonization 
formalism for a curved Fermi surface\cite{Kopietz96} to
gain insight into this paradox
between experiment and lowest order
perturbation theory. 
Our approach is most accurate for $q_c / k_F \equiv g \ll 1$, 
and resums self-energy and vertex corrections in a systematic way
to all orders in perturbation theory.
Note that for Fermi surfaces with constant curvature
it is not necessary to introduce the geometric patching
construction used in earlier formulations
of higher-dimensional bosonization\cite{Haldane92}.
Within the Gaussian approximation 
one obtains the following expression for the
Matsubara Green's function $G^{\alpha} ( \tilde{k} ) \equiv
G ( {\bf{k}}^{\alpha} + {\bf{k}} , i \tilde{\omega}_n )$,
 \begin{equation}
 G^{\alpha} ( \tilde{k} ) = \int d {\bf{r}} \int_{0}^{\beta} d \tau
 e^{-i ( {\bf{k}} \cdot {\bf{r}} -  \tilde{\omega}_n \tau )}
 G^{\alpha} ( {\bf{r}} , \tau )
 \; \; \; ,
 \label{eq:GbosFourier}
 \end{equation}
where
 $G^{\alpha} ( {\bf{r}} , \tau ) = \tilde{G}^{\alpha} ( {\bf{r}} , \tau )
 e^{Q^{\alpha} ( {\bf{r}} , \tau )}$. 
The Debye-Waller factor is of the form
 $Q^{\alpha} ( {\bf{r}}  , \tau ) = R^{\alpha} -
 S^{\alpha} ( {\bf{r}}  , \tau )$, 
with $R^{\alpha} = S^{\alpha} (0,0)$ and
 \begin{equation}
 S^{\alpha} ( {\bf{r}}  , \tau )
 = \frac{1}{\beta {{V}}} \sum_{\tilde{q}}
 \frac{ h^{{\rm RPA},\alpha}_{\tilde{q}}
  \cos ( {\bf{q}} \cdot {\bf{r}} - \omega_n \tau )  }
{ [ i \omega_n - \xi^{\alpha}_{\bf{q}}  ][ i \omega_n + \xi^{\alpha}_{ - {\bf{q}}} ] }
 \label{eq:Debye}
 \; \; \; .
 \end{equation}
The prefactor Green's function $\tilde{G}^{\alpha} ( {\bf{r}} , \tau )$ is 
 \begin{equation}
 \tilde{G}^{\alpha} ( {\bf{r}} , \tau ) = \frac{1}{\beta V}
 \sum_{\tilde{k}} e^{i ( {\bf{k}} \cdot {\bf{r}} - \tilde{\omega}_n \tau )}
 \frac{1 + Y^{\alpha} ( \tilde{k} )}{ i \tilde{\omega}_n 
 - \xi^{\alpha}_{ {\bf{k}} } - \Sigma_1^{\alpha}
 ( \tilde{k} ) }
 \label{eq:preffourier}
 \; \; \; ,
 \end{equation}
with the prefactor self-energy  
 \begin{eqnarray}
 {\Sigma}^{\alpha}_1 ( \tilde{k} )
 & = & 
  - \frac{1}{\beta V} \sum_{\tilde{q}}
 h^{{\rm RPA},\alpha}_{\tilde{q}}
 G^{\alpha}_1 ( \tilde{k} + \tilde{q} )
 \nonumber
 \\
 & \times &
 \frac{ 
  ( {\bf{k}} \cdot {\bf{q}} ) {\bf{q}}^2
  +
 \left( {\bf{k}} \cdot {\bf{q}} \right)^2 
  }
 { m^2 
[ i \omega_{n} - \xi^{\alpha}_{\bf{q}}  ]
[ i \omega_{n} + \xi^{\alpha}_{ - {\bf{q}}} ] }
 \; \; \; ,
 \label{eq:sigma1res}
 \end{eqnarray}
and the vertex function 
 \begin{eqnarray}
 Y^{\alpha} ( \tilde{k} ) & = &  
  \frac{1}{\beta {{V}}} \sum_{\tilde{q}}
 h^{{\rm RPA},\alpha}_{\tilde{q}}
 G^{\alpha}_1 ( \tilde{k} + \tilde{q})
 \nonumber
 \\
 & \times &
 \frac{ 
  {\bf{q}}^2 + 
 2 {\bf{k}} \cdot {\bf{q}}  }
 {
m [ i \omega_{n} - \xi^{\alpha}_{\bf{q}}  ]
[ i \omega_{n} + \xi^{\alpha}_{ - {\bf{q}}} ] }
 \; \; \; .
 \label{eq:Yres}
 \end{eqnarray}
Here 
$G_1^{\alpha} ( \tilde{k} )$ is related to the self-energy
in Eq.(\ref{eq:sigma1res}) via the Dyson equation
 $[ G^{\alpha}_1 ( \tilde{k} ) ]^{-1} =
 [ G^{\alpha}_0 ( \tilde{k} ) ]^{-1} 
 - \Sigma^{\alpha}_{1} ( \tilde{k} ) $.
To leading order in $g$ we may replace
$G_1^{\alpha}  \rightarrow
G_0^{\alpha} $ on the right-hand sides of 
Eqs.(\ref{eq:sigma1res}) and (\ref{eq:Yres}).
We shall further comment on this approximation below.
Note that $\Sigma_1^{\alpha}$ and
$Y^{\alpha}$ vanish
for linearized energy dispersion, i.e., for
$1/m = 0$.  
If we expand
Eq.(\ref{eq:GbosFourier})
to first order in $h^{ {\rm RPA} , \alpha}_{\tilde{q}}$,
we {\it{exactly}} recover
the leading term in perturbation theory\cite{Kopietz96}.

Consider first the constant part $R^{\alpha} $ of the Debye-Waller factor.
The integrations can be done with the same strategy as above:
we first perform the angular integration exactly,
approximate the result in the regime $ | \omega | 
{ \raisebox{-0.5ex}{$\; \stackrel{<}{\sim} \;$}} v_F q $, and then
perform the frequency integration.
For $\eta > 1$ we obtain
 $R^{\alpha} 
 = - r_{\eta} g $, where $r_{\eta} $ is
a positive constant of the order of unity.
Because by assumption
$g \ll 1$, we conclude
that $ R^{\alpha} $ is finite and small.
Obviously the extra factor
of $\cos ( {\bf{q}} \cdot {\bf{r}} - \omega_n \tau )$ 
in the expression for
$S^{\alpha} ( {\bf{r}} , \tau )$ given in Eq.(\ref{eq:Debye})
cannot lead to new divergencies. 
In fact, according to the  Fourier integral theorem 
the existence of $S^{\alpha} ( 0 , 0 )$ implies that
$S^{\alpha} ( {\bf{r}} , \tau )$ vanishes for
large ${\bf{r}}$ or $\tau$.
Thus, the factor $e^{Q^{\alpha} ( {\bf{r}} , \tau )}$ 
is always bounded and can be replaced by $e^{R^{\alpha}}$
in the large-distance and long-time limit.
If we had linearized the energy dispersion
in Eq.(\ref{eq:Debye}), 
we would have obtained a Debye-Waller factor
which diverges for large
${\bf{r}}$ or $\tau$\cite{Kwon94}.
Thus, the curvature of the Fermi surface 
is relevant in the sense that
the asymptotic Ward-identity\cite{Castellani94} leading to an exponentiation
of the perturbation series for the real space Green's function
cannot be used to resum the dominant singularities.
This is a consequence of the transversality of the gauge fields.%\cite{Castellani94b}.

\begin{figure}
%\vspace{1cm}
\epsfysize6cm 
\hspace{2cm}
\epsfbox{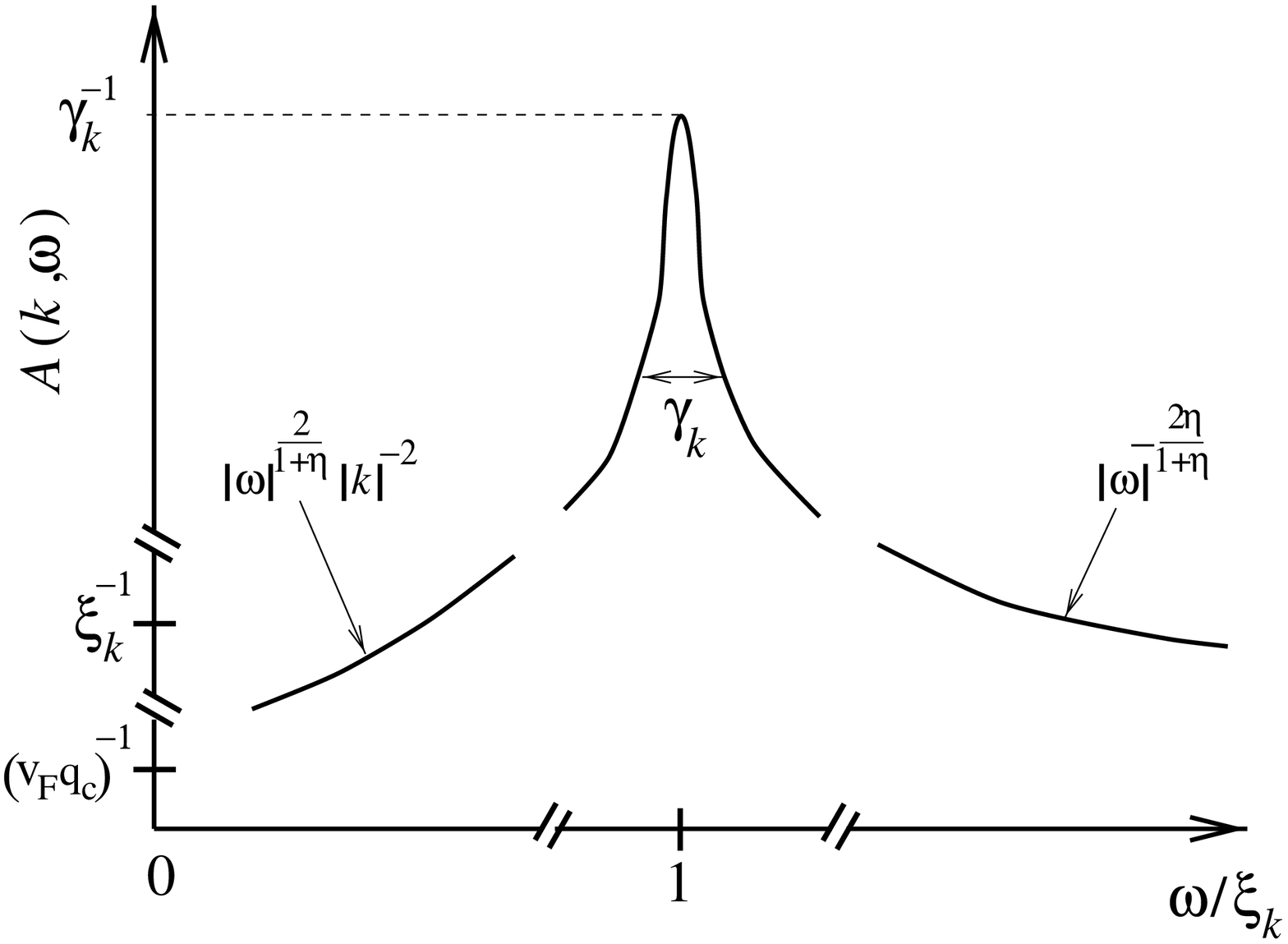}
%\vspace{1cm}
\caption{
Qualitative behavior of the spectral function 
for small $\bar{k}_{\|}$ and $\bar{\omega}$.
The subscript of $k_{\|}$ is omitted, i.e. $k$ stands
for $| {\bf{k}} | - k_F$.
Only the leading power laws are shown and
logarithmic corrections are ignored. The 
quasi-particle damping $\gamma_k$ is given in Eq.(\ref{eq:damping}).
The curves with the arrows indicate the
leading behavior for
${\omega} \ll  \xi_{{k}}$ and
${\omega} \gg  \xi_{{k}}$.
}
\label{fig:specfuncres}
\end{figure}

From Eqs.(\ref{eq:GbosFourier}) and (\ref{eq:preffourier})
we thus conclude that within higher-dimensional bosonization
the Green's function is 
 \begin{equation}
 {G}^{\alpha} ( \tilde{k} )  \approx
 \frac{ e^{R^{\alpha}} \left[
 1 + Y^{\alpha} ( \tilde{k} ) \right] }{ i \tilde{\omega}_n 
 - \xi^{\alpha}_{ {\bf{k}} } - \Sigma_1^{\alpha}
 ( \tilde{k} ) }
 \label{eq:preffinal}
 \; \; \; .
 \end{equation}
The most singular contributions to $\Sigma_1^{\alpha} ( \tilde{k} )$
and $Y^{\alpha} ( \tilde{k} )$
come from the regime of 
large wave-vectors $q 
{ \raisebox{-0.5ex}{$\; \stackrel{>}{\sim} \;$}}  | k_0 |$ and
small frequencies $| \omega |
{ \raisebox{-0.5ex}{$\; \stackrel{<}{\sim} \;$}}  | \tilde{\omega}_n |$.
So, the necessary integrations can be done in precisely the same manner as
in the case of $\Sigma_{\rm GW}^{\alpha}$ discussed above.
Retaining only the dominant terms for small $\bar{k}_{\|}$ and $\bar{\omega}_n$,
we find after some tedious but straightforward  manipulations
 \begin{eqnarray}
 \Sigma_{1}^{\alpha} ( k_{\|} , i \tilde{\omega}_n )
 & \approx  &
 - \frac{  i {\rm sgn} ( \tilde{\omega}_n ) }{ \pi m }
 \frac{ k_{\|}}{k_F}
 \int_{k_0}^{\kappa} dq   q
 \ln \left( 1 + \frac{ | \tilde{\omega}_n |}{ \Gamma_q } \right)
 \nonumber
 \\
 &  & \times
 \frac{ \frac{ q^2}{m} 
 }{G_0^{-1} - \frac{q^{2}}{m}  -  i \Gamma_q
 {\rm sgn} ( \tilde{\omega}_n )
 \frac{k_{\|}}{k_F} 
 }
 \nonumber
 \\
 &  & \times
 \frac{  G_0^{-1} - \left[
 \frac{q^2}{2 m} - i \Gamma_q {\rm sgn} ( \tilde{\omega}_n )  
 \right]}{   
 G_0^{-1} + \frac{k_{\|}}{k_F}
 \left[  \frac{q^2}{2 m} - i \Gamma_q {\rm sgn} ( \tilde{\omega}_n ) \right] }
 \label{eq:omegadone}
 \; \; \; ,
 \end{eqnarray}
 \begin{eqnarray}
 Y^{\alpha} ( k_{\|} , i \tilde{\omega}_n )
 & \approx &
  \frac{  i {\rm sgn} ( \tilde{\omega}_n ) }{ \pi m }
 \int_{k_0}^{\kappa} dq   q
 \ln \left( 1 + \frac{ | \tilde{\omega}_n |}{ \Gamma_q } \right)
 \nonumber
 \\
 &  & \times
 \frac{ \frac{ q^2}{m} 
 }{G_0^{-1} - \frac{q^{2}}{m}  -  i \Gamma_q
 {\rm sgn} ( \tilde{\omega}_n )
 \frac{k_{\|}}{k_F} 
 }
 \nonumber
 \\
 &  & \times
 \frac{ 1 }{   
 G_0^{-1} + \frac{k_{\|}}{k_F}
 \left[  \frac{q^2}{2 m} - i \Gamma_q {\rm sgn} ( \tilde{\omega}_n ) \right] }
 \label{eq:Yadone}
 \; \; \; .
 \end{eqnarray}
Having performed the frequency integrations, we may now analytically
continue these expressions to the real frequency axis.
Defining
 $\Sigma_1^{\alpha} ( k_{\|} , \omega + i 0^{+} )
  =  \Sigma^{\prime}_1  + i \Sigma^{\prime \prime}_1
  $ and
 $Y^{\alpha} ( k_{\|} , \omega + i 0^{+} )
  =  Y^{\prime} + i Y^{\prime \prime}$, the spectral function is
according to Eq.(\ref{eq:preffinal}) given by
 \begin{equation}
 A ( k_{\|} , \omega ) =  - \frac{e^{R^{\alpha}}}{ \pi} 
 \left[
 \frac{ 
 ( 1 + Y^{\prime} ) \Sigma^{\prime \prime}_1 + 
 Y^{\prime \prime}  \left(  \omega - \xi_{k_{\|}} - \Sigma^{\prime}_1 \right)}{
 ( \omega - \xi_{k_{\|}} - \Sigma^{\prime}_1 )^2 + ( \Sigma^{\prime \prime}_1 )^2 }
 \right]
 \label{eq:specbosform}
 \; \; .
 \end{equation}
We are interested in the behavior of
$A ( k_{\|}  , \omega )$ for small $ \bar{k}_{\|} = k_{\|} / q_c$ and
$\bar{\omega} = \omega / ( v_F q_c )$, but for arbitrary 
$\bar{\omega} / \bar{k}_{\|}$. 
Although
the precise form of 
$A ( k_{\|}  , \omega )$ can only be calculated numerically\cite{Kopietz97},
the qualitative behavior can be inferred from Eqs.(\ref{eq:omegadone}) and
(\ref{eq:Yadone}),
and is shown in Fig.\ref{fig:specfuncres}.
We first note that on resonance (i.e. for $\omega = \xi_{k_{\|}}$)
the leading behavior of $\Sigma_1^{\alpha} ( k_{\|} , 
\xi_{k_{\|}} + i 0^{+} )$ {\it{agrees exactly}} with the
perturbative self-energy
$\Sigma_{\rm GW}^{\alpha} ( k_{\|} , 
\xi_{k_{\|}} + i 0^{+} )$ discussed above.
This is due to the fact that on resonance
$G_0^{-1} = 0$, so that
the last factor in Eq.(\ref{eq:omegadone})
is proportional to $k_F / k_{\|}$ and cancels the 
small prefactor $k_{\|} / k_F$.
For the same reason we obtain from Eq.(\ref{eq:Yadone})
for $\bar{k}_{\|} \rightarrow 0$
 \begin{equation}
 Y^{\alpha} ( k_{\|} , \xi_{k_{\|}} + i 0^{+} )
 =   \frac{4 k_F \ln | \bar{k}_{\|} |  }{
    | {k}_{\|} | ( 1 + \eta ) }
  \left[ \frac{\eta}{4} - \frac{i {\rm{sgn}} ( \bar{k}_{\|} )}{\pi}
  \ln | \bar{k}_{\|} | \right]
 \label{eq:Yresonance}
 \; .
 \end{equation}
The important point is that on resonance this vertex
leads to a drastic
enhancement ($ \propto k_F / | k_{\|} |$) 
of the spectral weight,
and thus produces the quasi-particle peak in the spectral function shown in
Fig.\ref{fig:specfuncres}.
The width $\gamma_{k_{\|}}$ of the peak can be estimated from
the condition that the
term $G_0^{-1}$ in the denominator of the
last factor in Eqs.(\ref{eq:omegadone}) and (\ref{eq:Yadone})
becomes comparable with  
$\frac{k_{\|}}{k_F} \frac{q^2}{2m} $.
Keeping in mind that $q^2$ scales as $ | \bar{\omega}_n |^{\frac{2}{1 + \eta} }$,
it is easy to see that this leads to
 $| \bar{\omega} - \bar{\xi}_{k_{\|}} | 
 { \raisebox{-0.5ex}{$\; \stackrel{<}{\sim} \;$}}
 g^2 | \bar{k}_{\|} |^{ 1 + \frac{2}{1 + \eta}} $,
so that
 \begin{equation}
 {{\gamma}_{k_{\|}}}/ {{\xi}_{k_{\|}}}
 \approx  g^2 \left| \bar{k}_{\|} \right|^{\frac{2}{1 + \eta}}
 = ( q_c / k_F )^{\frac{2 \eta }{1 + \eta}} ( | k_{\|} | / k_F )^{\frac{2}{1 + \eta}}
  \label{eq:damping}
  \; \; \; .
  \end{equation}
Hence, the width of the peak vanishes faster than
$\xi_{k_{\|}}$ as
$k_{\|} \rightarrow 0$. We conclude that
{\it{the quasi-particle is well-defined and has the same energy dispersion as the bare
particle}}. Thus, the effective mass $m^{\ast}$ of the
composite fermions can be identified with the mean-field mass $m$\cite{footnote2}.
Note, however, that $\gamma_{k_{\|}}$ vanishes slower than $k_{\|}^2$, so that
the damping is anomalously large. 
Off resonance the self-energy contribution $\Sigma^{\alpha}_1$ is completely
negligible. Then
the behavior of the spectral function is determined
by the imaginary part $Y^{\prime \prime}$ of the vertex function. This is shown in 
Fig.\ref{fig:specfuncres}.
Obviously there is considerable weight off resonance, and the
overall shape of the spectral function cannot be approximated by a Lorentzian.
More accurate numerical results will be presented elsewhere\cite{Kopietz97}.

Finally, let us justify the replacement $G_1^{\alpha} \rightarrow G_0^{\alpha}$
on the right-hand sides of Eqs.(\ref{eq:sigma1res}) and (\ref{eq:Yres}).
Because the integrals are dominated by the regime 
$q 
{ \raisebox{-0.5ex}{$\; \stackrel{>}{\sim} \;$}} k_0$ and
$| \omega | 
{ \raisebox{-0.5ex}{$\; \stackrel{<}{\sim} \;$}} | \tilde{\omega}_n |$,
Eqs.(\ref{eq:sigma1res}) and (\ref{eq:Yres}) 
are determined by $G_1^{\alpha} ( \tilde{k} + \tilde{q} )$
off resonance. Recall that in this regime
$\Sigma_1^{\alpha}$ is negligible.

In summary, using a controlled  non-perturbative
approach we have shown that the 
quasi-particle picture
for composite fermions in the half-filled Landau level
remains valid even if the
infrared fluctuations of the Chern-Simons gauge field
are taken into account,
in agreement with the experimental fact\cite{Willett93} that
composite fermions behave like non-interacting particles.
%that the single-particle picture works remarkably well.
Moreover, the infrared fluctuations 
do not lead to a singularity in the effective mass $m^{\ast}$.
Thus, the experimentally observed 
enhancement of $m^{\ast}$ must have a different origin.
We speculate that this is related to
Landau level mixing or short-wavelength fluctuations,
both of which are not correctly described 
by the Chern-Simons approach\cite{Halperin93}. 
Indeed, there exists some experimental evidence
for a correlation between
Landau level mixing and an enhanced $m^{\ast}$: 
in the experiment by Manoharan {\it{et al.}}\cite{Willett93},
where the Landau level mixing was  particularly strong, the
increase of $m^{\ast}$ close to $\nu = \frac{1}{2}$
was larger than in the other experiments\cite{Willett93}.

We are indebted to A. Castro-Neto for discussions
and to K. Sch\"{o}nhammer for comments on the manuscript.
This work was supported in part (GC) by the Department of Energy contract
No. DE-AC02-76CH00016. 
%
%              R E F E R E N C E S
%


\begin{thebibliography}{99}
\bibitem{footnoteeffmess}
The effective mass can be defined  by
writing the energy dispersion for momenta ${\bf{k}}$ close to the
Fermi surface as
$\frac{{k}_F}{m^{\ast}}
( | {\bf{k}} | - k_F )$. Hence, a diverging $m^{\ast}$ means that
the dispersion vanishes faster than linear as $| {\bf{k}} | \rightarrow k_F$.
\bibitem{footnote2}
In deriving Eq.(\ref{eq:omegadone}) we have ignored sub-leading terms
which possibly yield a finite correction to $m$ proportional to $g$.
This is justified because by assumption $g \ll 1$.
\end{thebibliography}
\end{document}